\documentclass[english]{article}
\usepackage[T1]{fontenc}
\usepackage[latin9]{inputenc}
\usepackage{geometry}
\usepackage{color}
\usepackage{float}
\usepackage{amsmath}
\usepackage{graphicx}

\makeatletter

\newcommand*\LyXThinSpace{\,\hspace{0pt}}
\providecommand{\tabularnewline}{\\}

\newcommand{\lyxaddress}[1]{
\par {\raggedright #1
\vspace{1.4em}
\noindent\par}
}

\@ifundefined{showcaptionsetup}{}{%
 \PassOptionsToPackage{caption=false}{subfig}}
\usepackage{subfig}
\makeatother

\usepackage{babel}
\begin{document}

\title{Mathematical Model of ingested glucose in Glucose-Insulin Regulation }

\author{Sourav Chowdhury\thanks{Short term project student working under Dr. Suparna Roychowdhury
and Dr. Indranath Chaudhuri, Department of Physics, St. Xavier's College
(Autonomous), Kolkata}, Sourabh Kumar Manna\thanks{Short term project student working under Dr. Suparna Roychowdhury
and Dr. Indranath Chaudhuri, Department of Physics, St. Xavier's College
(Autonomous), Kolkata}, Suparna Roychowdhury, Indranath Chaudhuri}
\maketitle

\lyxaddress{\begin{center}
Department of Physics, St. Xavier's College(Autonomous), 30 Mother
Teresa Sarani, Kolkata-700016.
\par\end{center}}
\begin{abstract}
Here, we develop a mathematical model for glucose-insulin regulatory
system. The model includes a new parameter which is the amount of
ingested glucose. Ingested glucose is an external glucose source coming
from digested food. We assume that the external glucose or ingested
glucose decays exponentially with time. We establish a system of three
linear ordinary differential equations with this new parameter, derive
stability analysis and the solution of this model. 
\end{abstract}

\section*{{\large{}Keywords }}

Mathematical model, Diabetes mellitus, Linear system, Ingested glucose,
Glucose tolerance test, Natural time period, Stability analysis.

\section{Introduction}

Glucose which we get from food is very important for the human body
because it is like  fuel and energy source for cells as well as the
human body. However diabetes is a condition when blood glucose level
exceeds the normal range (75 -110 mg/dl) for a long period of time.
In 2017, 4 million people died due to diabetes and approximately 425
million adults in the world had diabetes. Statistics says that by
2045 the number of people with diabetes will rise to 629 million.
More than 1,106,500 children have type 1 diabetes and more than 21
million births are affected by diabetes during pregnancy (2017). Around
352 million people have risk to develop type 2 diabetes \cite{key-1}.
India was ranked up from 11th (2005) to 7th (2016) due to number of
deaths by diabetes and there are 70 million people who suffered from
diabetes and statistics says these numbers will grow more than double
in the next decade \cite{key-2}. For normal person blood glucose
level reaches to a homeostasis with the help of two types of hormones,
first which reduces blood glucose level like Insulin, Amylin, Somatostatin
and second which raises blood glucose level like Glucagon, Epinephrine
(adrenaline), Growth Hormone, Thyroxine \cite{key-3}. However when
there is a disorder in secretion of Insulin or cells becomes resistant
of Insulin or both, then lower amount of glucose reaches to the cells
and blood glucose level is very high than fasting blood glucose level
for a long period of time. Since the lower amount of glucose reaches
the cells, body weakens and person become diabetic \cite{key-4}. 

There are mainly three types of diabetes :
\begin{itemize}
\item \textbf{Type 1 diabetes mellitus} - Type 1 diabetes mellitus previously
known as ``insulin-dependent diabetes mellitus'' (IDDM) or ``juvenile
diabetes''. Almost 10\% of worldwide diabetics have type 1 diabetes
and most of them are children. This type of diabetes caused due to
deficiency of insulin in blood. Feeling very thirsty, urinating frequently,
feeling very tired, weight loss, constant hunger are the main symptoms
of type 1 diabetes. \cite{key-5,key-6,key-7}
\item \textbf{Type 2 diabetes mellitus - }Type 2 diabetes mellitus previously
known as ``non-insulin-dependent diabetes mellitus'' (NIDDM) or
``adult onset diabetes''. Almost 90\% of worldwide diabetics have
type 2 diabetes and most of them are adults. This type of diabetes
is caused due to the insulin resistance in the cells of the body (may
be due to problem in insulin receptors in the cells). Symptoms are
almost similar as type 1 diabetes. However people who suffers from
type 2 diabetes are often obese. \cite{key-5,key-6,key-7}
\item \textbf{Gestational diabetes - }Gestational diabetes is a type of
diabetes which develops in some women when they are pregnant. Most
of the time, this type of diabetes goes away after the baby is born.
However they and their children have increased risk of type 2 diabetes.
Gestational diabetes is mainly diagnosed through parental screening
rather than through symptoms. \cite{key-5,key-6,key-7}
\end{itemize}
Mathematical models are very important to understand the dynamic behavior
of complex biological systems. There are various mathematical models,
statistical methods and algorithms to understand different aspects
of diabetes. Models can be classified between two categories : 
\begin{itemize}
\item \textbf{Clinical models} - Clinical models are developed to understand
a disease more accurately so that we can find a better cure of it.
There are various existing models for diagnostics like GTT (glucose
tolerance test), IVGTT (intravanous glucose tolerance test), OGTT
(oral glucose tolerance test), FSIVGTT (frequently sampled intravenous
glucose tolerance test). \cite{key-4,key-8,key-9}
\item \textbf{Non-clinical models - }Non-clinical models are developed from
the partial knowledge of the system. There are various non-clinical
models to understand insulin-glucose dynamics and also there are many
different types of non-clinical models. \cite{key-4,key-8,key-9}
\end{itemize}
Since diabetes is very complex in nature, these models needs to be
upgraded with respect to the experimental knowledge. Here we present
a realistic model by considering a new parameter which represents
ingested glucose (External glucose which is acquired from intake food).

The paper is arranged in the following way : importance of glucose
and insulin and their role in the human body, discussion of the previous
model (briefly) and the new model, analysis of stability and calculation
for model, model fitting of a data-set and finally concluding remarks. 

\section{Glucose-Insulin dynamics in the human body}

In this section we will discuss briefly, how glucose-insulin plays
an important role in the human body. We start with a simple block
diagram which represents importance of glucose in the human body.

\begin{figure}[H]
\begin{centering}
\includegraphics[scale=0.7]{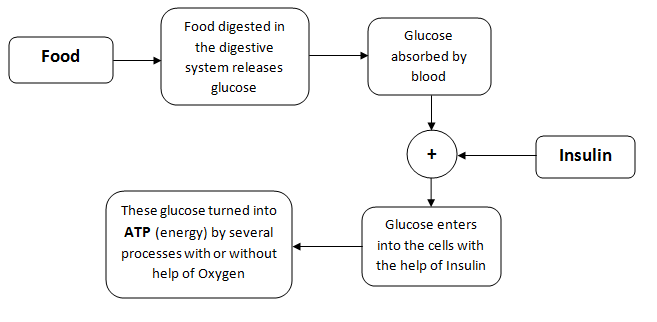}
\par\end{centering}
\caption{Block diagram to show how glucose is converted to energy.}
\end{figure}

Glucose is converted to energy (ATP) in the cells by glycolysis and
other processes. In some cells glucose turns into energy with the
help of oxygen and in some cells it turns into energy without oxygen
\cite{key-10}. Also insulin is very important, because with the help
of insulin glucose can enter into cells. If there is less insulin
or the body cells are insulin resistant then glucose cannot enter
into the cells, and it remains in the blood creating different complications
\cite{key-3,key-6}.

\begin{figure}[H]
\begin{centering}
\includegraphics[scale=0.7]{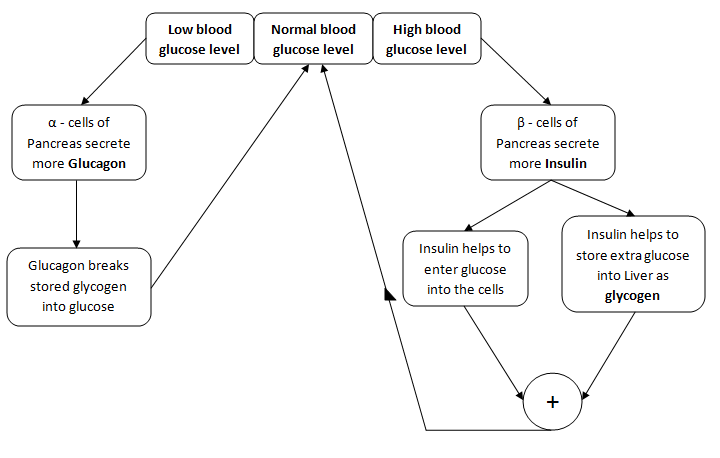}
\par\end{centering}
\caption{Block diagram to show how homeostasis is maintained for a normal person.}
\end{figure}

The human body needs to maintain homeostasis (dynamic equilibrium
in human body) of blood glucose level. To maintain this homeostasis,
mainly insulin and glucagon work together. Insulin is released from
the $\beta$-cells of pancreas, and glucagon is released from the
$\alpha$-cells of pancreas. Figure 2 shows how human body maintains
homeostasis with the help of these two hormones. When blood glucose
level of the human body is higher than the normal level, the pancreas
secretes more insulin. Insulin helps glucose to enter the cells. This
also helps the extra glucose to get stored into the liver as glycogen,
thus reducing the blood glucose level to maintain the homeostasis.
However, when blood glucose level is low, the pancreas secretes glucagon
which breaks stored glycogen into glucose and increases blood glucose
level to normal level \cite{key-4,key-7,key-11}.

\begin{figure}[H]
\begin{centering}
\includegraphics[scale=0.7]{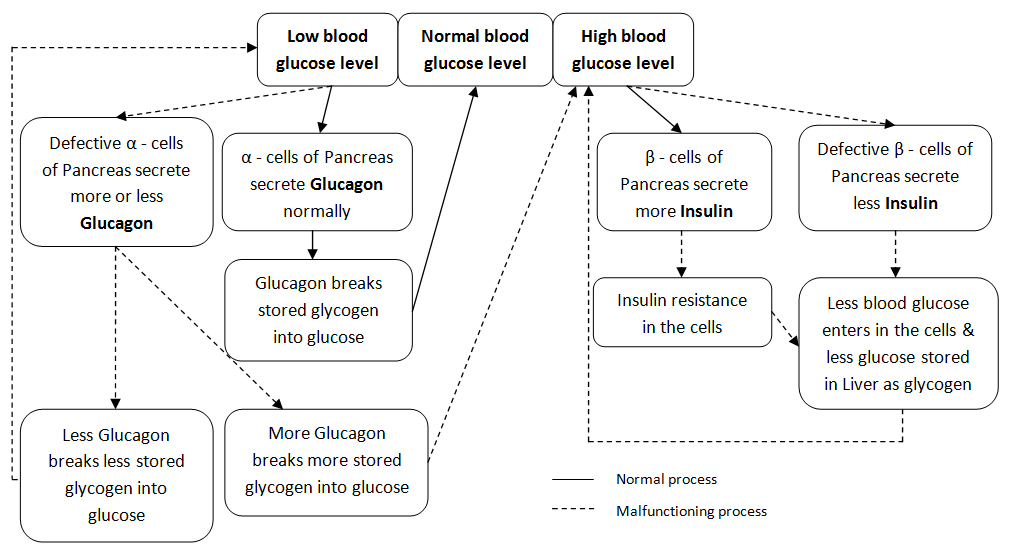}
\par\end{centering}
\caption{Block diagram to show how glucose level varies for diabetic person.}
\end{figure}

However for a diabetic person, the human body cannot maintain homeostasis
of blood glucose level {[}Figure 3{]}. In type-1 diabetes, $\beta$-cells
of pancreas are destroyed by ones own immune system (T-cells). Thus
less insulin is produced and for this insulin deficiency blood glucose
level is high for a very long period. 

In type-2 diabetes, cells of the human body becomes insulin resistant.
So, insufficient amount of blood glucose enters into the cells. Also
low amount of blood glucose gets stored into liver in the form of
glycogen. To overcome this situation pancreas secretes excess insulin.
If this excess insulin is enough to bring back blood glucose level
to the normal level then the person is pre-diabetic. However if this
excess insulin is not enough then the person is type-2 diabetic. Other
reason for which blood glucose level can be high is due to malfunctioning
$\alpha$-cells of pancreas. In this case they produce excess amount
of glucagon \cite{key-4,key-6,key-7}.

\section{Modeling glucose-insulin dynamics}

In this section we first discuss a model suggested by previous authors.
Next we go on to discuss the relevant factors that we have introduced
into our model.

In 1964, E. Ackerman, J. W. Rosevear and W. F. McGuckin developed
a mathematical model of Glucose-tolerance test \cite{key-12}. In
this model they have considered $G(t)$ as blood-glucose concentration
and $H(t)$ as blood-insulin concentration. Here $G_{b}$ and $H_{b}$
were defined as fasting blood glucose level and fasting hormone (insulin)
level respectively. $D_{2}(t)$ and $h(t)$ were defined as difference
of the blood-glucose level and blood-hormone (insulin) level from
the fasting levels respectively. So, $D_{2}(t)=G(t)-G_{b}$ and $h(t)=H(t)-H_{b}$.
The main equations used in that model are,

\begin{alignat*}{1}
\frac{dD_{2}}{dt}= & \;-a_{1}D_{2}-a_{2}h+I,\\
\frac{dh}{dt}= & \;b_{1}D_{2}-b_{2}h.
\end{alignat*}

Where, 

$a_{1}=$Rate constant of glucose removal independent of insulin.

$a_{2}=$Rate constant of glucose removal dependent of insulin.

$b_{1}=$Rate constant of insulin release due to glucose.

$b_{2}=$Rate constant of insulin removal independent of glucose.

$I(t)$= Rate of increase of blood glucose due to absorption of glucose
from intestines.

We have modified the above model taking into account the contribution
of the ingested glucose in the form of an additional differential
equation. We explain this model using a block diagram where various
parts of the human body represented by compartments. {[}Figure 4{]} 

\begin{figure}[H]
\begin{centering}
\includegraphics[scale=0.6]{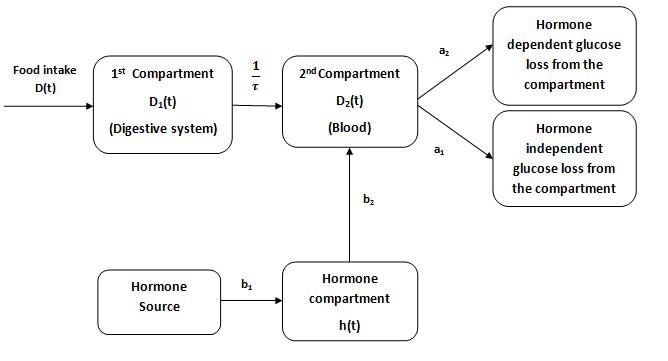}
\par\end{centering}
\caption{Diagram of simplified model of blood glucose regulatory system.}
\end{figure}

Let at time $t$ glucose disturbance of the digestive system (1st
compartment) is $D_{1}(t)$. In this system we assume that the fasting
glucose level is zero, thus glucose level and glucose disturbance
is same. Let at time $t$ the blood (2nd compartment) glucose level
is $G(t)$ and the effective hormone level is $H(t)$. By the term
effective hormone level means the net effect of all hormones (like
insulin, glucagon) which can increase or decrease blood glucose level.
The fasting blood glucose and the effective hormone levels $G_{b}$
and $H_{b}$ respectively. Let, $D_{2}(t)$ and $h(t)$ are the disturbances
of the blood glucose level and the effective hormone level at time
$t$ respectively. Thus we can write $D_{2}(t)=G(t)-G_{b}$ and $h(t)=H(t)-H_{b}$.
\cite{key-12,key-13}

In this model we make following assumptions :
\begin{itemize}
\item Rate of decrease in glucose disturbance in digestive system is proportional
to its glucose disturbance ($D_{1}(t)$) at time $t$.
\item Rate of increase in glucose disturbance in blood is proportional to
the glucose which enters blood from digestive system ($D_{1}(t)$)
at time $t$ .
\item Rate of hormone independent decrease of glucose disturbance in blood
is proportional to its glucose disturbance ($D_{2}(t)$) at time $t$
.
\item Rate of hormone dependent decrease of glucose disturbance in blood
is proportional to the effective hormone disturbance ($h(t)$) at
time $t$.
\item Rate of increase in effective hormone disturbance is proportional
to the glucose disturbance in blood ($D_{2}(t)$) at time $t$.
\item Rate of decrease in effective hormone disturbance is proportional
to the effective hormone disturbance ($h(t)$) at time $t$.
\end{itemize}
Using these assumptions the model equations can be written as,

\begin{alignat}{2}
\frac{dD_{1}}{dt}= & \;-\frac{D_{1}}{\tau}, & \qquad D_{1}(0)=A_{G}D,\\
\frac{dD_{2}}{dt}= & \;\frac{D_{1}}{\tau}-a_{1}D_{2}-a_{2}h, & \qquad D_{2}(0)=0,\\
\frac{dh}{dt}= & \;b_{1}D_{2}-b_{2}h, & \qquad h(0)=0.
\end{alignat}

$\tau=$Time constant of decreasing glucose level in the digestive
system, which is the total time to decrease glucose level to $1/e$
of the maximum value.

$a_{1}=$Rate constant of the hormone independent decrease of glucose
level in the blood.

$a_{2}=$Rate constant of the hormone dependent decrease of glucose
level in the blood.

$b_{1}=$Rate constant of release of the hormone due to blood glucose
disturbance. 

$b_{2}=$Rate constant for the removal of the hormone due to disturbance
of the blood hormone level.

$D=$Amount of food that has been taken.

$A_{G}=$Percentage of glucose obtained in the body from the food
that has been taken.

\section{Stability analysis of this model}

To determine fixed points we can write,

\begin{alignat}{1}
\frac{dD_{1}}{dt}= & \;\frac{dD_{2}}{dt}=\frac{dh}{dt}=0.\nonumber \\
\intertext{\textrm{Substituting \ensuremath{{\displaystyle \mathit{\frac{dD_{\mathrm{1}}}{dt}=\mathrm{0}}}} in equation (1) we get,}}D_{1}= & \;0=D_{10}.\\
\intertext{\textrm{Now making \ensuremath{{\displaystyle \mathit{\frac{dD_{\mathrm{2}}}{dt}=\mathrm{0}}}} and \ensuremath{{\displaystyle \mathit{\frac{dh}{dt}=\mathrm{0}}}} and putting \ensuremath{{\displaystyle \mathit{D}_{\mathrm{10}}}} in equation (2) we get,}}\frac{D_{10}}{\tau}- & a_{1}D_{20}-a_{2}h_{0}=0,\nonumber \\
a_{1}D_{20} & +a_{2}h_{0}=0.\\
\intertext{\textrm{From equation (3) we get,}}b_{1}D_{20} & -b_{2}h_{0}=0.\\
\intertext{\textrm{Solving equation (5), (6) we get,}}D_{20}= & 0\\
\intertext{\textrm{and}}h_{0}= & 0.
\end{alignat}

So the fixed point is $\mathbf{\mathit{\mathrm{(D_{10},D_{20},h_{0})=(0,0,0)}}}$.
Thus equations $(1),\:(2)\:\textrm{and}\:(3)$ can be written in the
matrix form as,

\[
\left(\begin{array}{c}
\frac{dD_{1}}{dt}\\
\frac{dD_{2}}{dt}\\
\frac{dh}{dt}
\end{array}\right)=\left(\begin{array}{ccc}
-\frac{1}{\tau} & 0 & 0\\
\frac{1}{\tau} & -a_{1} & -a_{2}\\
0 & b_{1} & -b_{2}
\end{array}\right)\left(\begin{array}{c}
D_{1}\\
D_{2}\\
h
\end{array}\right)
\]

Let, 
\[
\left(\begin{array}{ccc}
-\frac{1}{\tau} & 0 & 0\\
\frac{1}{\tau} & -a_{1} & -a_{2}\\
0 & b_{1} & -b_{2}
\end{array}\right)=A
\]

Now characteristic equation is given by,

\[
det(A-\lambda I)=0
\]

Here $\lambda$ is the eigen values of matrix $A$. So,

\[
\left|\begin{array}{ccc}
(-\frac{1}{\tau}-\lambda) & 0 & 0\\
\frac{1}{\tau} & (-a_{1}-\lambda) & -a_{2}\\
0 & b_{1} & (-b_{2}-\lambda)
\end{array}\right|=0
\]

From this determinant we can write,

\[
\left(-\frac{1}{\tau}-\lambda\right)\left[\left(a_{1}+\lambda\right)\left(b_{2}+\lambda\right)+a_{2}b_{1}\right]=0.
\]

The above equation can write into two algebraic equations as, $\left(-\frac{1}{\tau}-\lambda\right)=0$
and $\lambda^{2}+(a_{1}+b_{2})\lambda+(a_{1}b_{2}+a_{2}b_{1})=0$.

Hence the first eigenvalue is $\lambda_{1}=-\frac{1}{\tau}$ where
$\tau>0$ so $\lambda_{1}<0$ and the other two eigenvalues are 

\[
\lambda_{2}=\frac{-(a_{1}+b_{2})+\sqrt{(a_{1}-b_{2})^{2}-4a_{2}b_{1}}}{2},
\]

\[
\lambda_{3}=\frac{-(a_{1}+b_{2})-\sqrt{(a_{1}-b_{2})^{2}-4a_{2}b_{1}}}{2}.
\]

If $(a_{1}-b_{2})^{2}>4a_{2}b_{1}$ then $\sqrt{(a_{1}-b_{2})^{2}-4a_{2}b_{1}}<(a_{1}+b_{2})$.
So, $\lambda_{2}$ and $\lambda_{3}$ are both negative.

If $(a_{1}-b_{2})^{2}<4a_{2}b_{1}$ then $\sqrt{(a_{1}-b_{2})^{2}-4a_{2}b_{1}}$
is a complex quantity. Hence $\lambda_{2}$ and $\lambda_{3}$ are
both complex with negative real part.

Hence the model is stable around $(0,0,0)$.

\vspace{0.5cm}

\begin{figure}[H]
\begin{centering}
\subfloat{\includegraphics[scale=0.65]{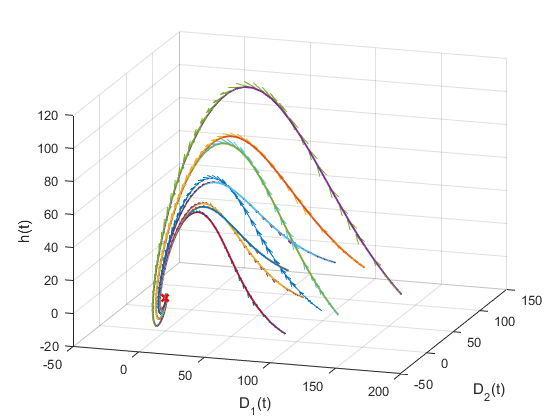}}
\par\end{centering}
\caption{Phase portraits of these model shows it is globally stable. (0,0,0)
point is represented in the plots as red \textcolor{black}{(}\textcolor{red}{$\mathbf{\times}$}).}
\end{figure}

Figure 5 represents a three dimensional phase portrait with different
initial conditions. From this plots we can see that all the trajectories
converge to the point (0,0,0). Hence (0,0,0) is a stable equilibrium
point and the model is globally stable.

\section{Calculation of glucose level in blood}

In this section we obtain a solution for blood glucose level from
the model equations by considering all initial conditions.

\noindent Solving equation $(1)$ we get,

\begin{alignat}{1}
D_{1}= & A_{0}e^{-\frac{t}{\tau}}.\nonumber \\
\intertext{\textrm{Where \ensuremath{\mathit{A}_{0}}= Integrating constant. Substituting initial conditions,}}D_{1}(0)= & A_{G}D=A_{0}.\nonumber \\
\intertext{\textrm{Hence,}}D_{1}(t)= & A_{G}De^{-\frac{t}{\tau}}.\\
\intertext{\textrm{ Differentiating equation (2),}}\frac{d^{2}D_{2}}{dt^{2}}= & -a_{1}\frac{dD_{2}}{dt}-a_{2}\frac{dh}{dt}+\frac{1}{\tau}\frac{dD_{1}}{dt}.\\
\intertext{\textrm{Substituting the relation of \ensuremath{\frac{\mathit{dh}}{\mathit{dt}}} from equation (3) in equation (10) we get,}}\frac{d^{2}D_{2}}{dt^{2}}= & -a_{1}\frac{dD_{2}}{dt}-a_{2}(b_{1}D_{2}-b_{2}h)+\frac{1}{\tau}\frac{dD_{1}}{dt}.\nonumber \\
\intertext{\textrm{After substituting relation of \ensuremath{{\displaystyle \mathit{h}}} from equation (2) in equation (11) and rearranging we get,}}\frac{d^{2}D_{2}}{dt^{2}}+ & (a_{1}+b_{2})\frac{dD_{2}}{dt}+(a_{1}b_{2}+a_{2}b_{1})D_{2}=\frac{b_{2}}{\tau}D_{1}+\frac{1}{\tau}\frac{dD_{1}}{dt}.\\
\intertext{\textrm{Now using, \ensuremath{{\displaystyle \mathit{a_{\mathrm{1}}+b_{\mathrm{2}}=\mathrm{\mathit{\mathrm{2}}}\alpha}}}, which is sum of the rate of hormone independent glucose removal and the hormone removal and \ensuremath{{\displaystyle \mathit{{\textstyle a_{\mathrm{1}}b_{\mathrm{2}}+a_{\mathrm{2}}b_{\mathrm{1}}=\omega_{\mathrm{0}}^{\mathrm{2}}}}}} we get,}}\frac{d^{2}D_{2}}{dt^{2}}+ & 2\alpha\frac{dD_{2}}{dt}+\omega_{0}^{2}D_{2}=\frac{b_{2}}{\tau}D_{1}+\frac{1}{\tau}\frac{dD_{1}}{dt}.\\
\intertext{\textrm{Substituting \ensuremath{{\displaystyle \mathit{D_{1}(t)}}} from equation (9) we get,}}\frac{d^{2}D_{2}}{dt^{2}}+ & 2\alpha\frac{dD_{2}}{dt}+\omega_{0}^{2}D_{2}=(\frac{b_{2}A_{G}D}{\tau}-\frac{A_{G}D}{\tau^{2}})e^{-\frac{t}{\tau}}.\\
\intertext{\textrm{Using \ensuremath{{\displaystyle {\displaystyle \mathit{C}\mathit{=\frac{b_{\mathrm{2}}A_{G}D}{\tau}-\frac{A_{G}D}{\tau^{\mathrm{2}}}}}}}}}\frac{d^{2}D_{2}}{dt^{2}}+ & 2\alpha\frac{dD_{2}}{dt}+\omega_{0}^{2}D_{2}=Ce^{-\frac{t}{\tau}}.\\
\intertext{\textrm{Solving equation (14),}}D_{2}= & e^{-\alpha t}[A_{1}cos(\omega t)+A_{2}sin(\omega t)]+C_{0}e^{-\frac{t}{\tau}}.
\end{alignat}

\noindent Where $\omega^{2}=\omega_{0}^{2}-\alpha^{2}$ and $\alpha^{2}<\omega_{0}^{2}$
and $C_{0}=\frac{A_{G}D(b_{2}\tau-1)}{(\alpha\tau-1)^{2}+\omega^{2}\tau^{2}}$.

\vspace{0.2cm}

\noindent Now applying initial conditions $D_{2}(0)=0$ and $\frac{dD_{2}}{dt}\mid_{t=0}=\frac{A_{G}D}{\tau}$
in equation $(15)$ we get,

\vspace{0.2cm}

\noindent So finally we get $A_{1}=-C_{0}$ and $A_{_{2}}=\frac{A_{G}D+C_{0}-\alpha C_{0}\tau}{\omega\tau}$.

\vspace{0.2cm}

\noindent So the disturbance in 2nd compartment is,

\begin{alignat}{1}
D_{2}= & e^{-\alpha t}[-C_{0}cos(\omega t)+\frac{A_{G}D+C_{0}-\alpha C_{0}\tau}{\omega\tau}sin(\omega t)]+C_{0}e^{-\frac{t}{\tau}}.\\
\intertext{\textrm{Hence glucose level in 2nd compartment is,}}G(t)= & G_{b}+D_{2}=G_{b}+e^{-\alpha t}[-C_{0}cos(\omega t)+\frac{A_{G}D+C_{0}-\alpha C_{0}\tau}{\omega\tau}sin(\omega t)]+C_{0}e^{-\frac{t}{\tau}}.
\end{alignat}

From equation $(16)$ we can see that the solution of $D_{2}(t)$
is like a oscillator with a small damping. $h(t)$ can have both positive
and negative values. Positive $h(t)$ corresponds to the effective
hormone level which decreases the blood glucose level. Similarly,
negative $h(t)$ corresponds to the effective hormone level which
increases the blood glucose level. As we see, $\alpha$ and $\omega$
can have many different values \cite{key-12,key-13}. So the crucial
parameter turns out to be $\omega_{0}$ which in other words is a
natural time period $T_{0}=\frac{2\pi}{\omega_{0}}$. If $T_{0}<4.0$
hours then the person is normal and if $T_{0}\geq4.0$ hours then
the person is pre-diabetic \cite{key-12,key-13}.

\section{Modeling with experimental data}

In this section, the blood glucose level $G(t)$ that is obtained
from our model is fitted with a data-set and essential parameters
of our model are estimated. The data-set represents a glucose vs time
data for a normal (Subject A) and a diabetic (Subject B) person. This
data-set is taken from ``Modeling Diabetes'' by Joseph M. Mahaffy
(Math 636 - Mathematical Modeling) \cite{key-16}. 

\begin{table}[H]
\centering{}%
\begin{tabular}{|c|c|c|}
\hline 
t in hour & Subject A & Subject B\tabularnewline
\hline 
\hline 
0 & 70 & 100\tabularnewline
\hline 
0.5 & 150 & 185\tabularnewline
\hline 
0.75 & 165 & 210\tabularnewline
\hline 
1 & 145 & 220\tabularnewline
\hline 
1.5 & 90 & 195\tabularnewline
\hline 
2 & 75 & 175\tabularnewline
\hline 
2.5 & 65 & 105\tabularnewline
\hline 
3 & 75 & 100\tabularnewline
\hline 
4 & 80 & 85\tabularnewline
\hline 
6 & 75 & 90\tabularnewline
\hline 
\end{tabular}\caption{Blood glucose level data for normal (Subject A) and diabetic person
(Subject B).}
\end{table}

\begin{figure}[H]
\centering{}\includegraphics[scale=0.75]{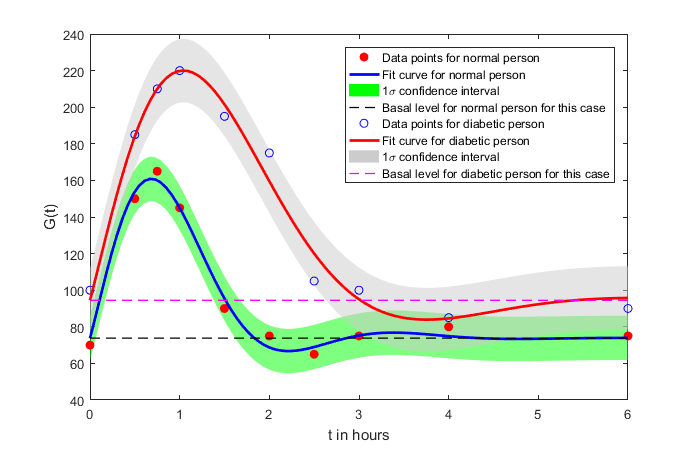}\caption{Fitted curve for normal and diabetic persons with $1\sigma$ interval.}
\end{figure}

Figure 6 shows the best fit curve with $1\sigma$ confidence interval.
$1\sigma$ confidence interval confidence interval is a statistically
calculated interval from a observed data-set. This interval tells
us that the true value of the parameter may be lie within it. Here,
1$\sigma$ confidence interval is defined as $G\left(t\right)\pm\frac{\sigma}{\sqrt{N}}$,
$\sigma$=standard deviation of data and $N=$total number of data.
If, $N$ increases then confidence interval decreases but confidence
remains same.

\pagebreak{}

\noindent The values of the obtained parameters are,

\begin{table}[H]
\begin{centering}
\begin{tabular}{|c||c|c|c|c|c|}
\hline 
Subject & $G_{b}$ & $\alpha$ & $\omega$ & $C_{0}$ & $\tau$ in hour\tabularnewline
\hline 
\hline 
A & 73.8161 & 1.1733 & 2.4128 & 116.4327 & 0.6447\tabularnewline
\hline 
B & 94.4838 & 0.8685 & 1.2823 & 208.3634 & 0.6242\tabularnewline
\hline 
\end{tabular}
\par\end{centering}
\caption{Values of the fit parameters}
\end{table}

\begin{table}[H]
\centering{}%
\begin{tabular}{|c||c|c|c|c|}
\hline 
Subject & $\alpha$ & $\omega$ & $\omega_{0}=\sqrt{\omega^{2}+\alpha^{2}}$ & $T_{0}=\frac{2\pi}{\omega_{0}}$ in hour\tabularnewline
\hline 
\hline 
A & 1.1733 & 2.4128 & 2.6829 & 2.3418\tabularnewline
\hline 
B & 0.8685 & 1.2823 & 1.5487 & 4.0569\tabularnewline
\hline 
\end{tabular}\caption{Values of main parameters $\omega_{0}$ and $T_{0}$}
\end{table}

$G_{b}$= The fasting blood glucose level. It can vary person to person.

$\alpha$= Decay parameter of the damped oscillator.

$\omega$= Angular frequency of damped oscillator.

From Table 2 we can see that value of both $\alpha$ and $\omega$
for diabetic person is less than the normal person.

$\tau$= Time constant of decreasing glucose level in the digestive
system, which is the total time to decrease glucose level to $1/e$
of the maximum value. From Table 2 we can see that the value of this
parameter is nearly equal for both normal and diabetic person. 

$\omega_{0}$= Natural angular frequency of damped oscillator.

$T_{0}$= Natural time period of damped oscillator. From Table 3 we
can see that value of $T_{0}$ for normal person is small than the
diabetic person.

Here we plot the data-set and fit the function with 1$\sigma$ confidence
interval {[}Figure 6{]}. Values of $T_{0}$ of Subject A and Subject
B are 2.3418 hours and 4.0569 hours respectively {[}from Table 3{]},
which indicates Subject A is normal and Subject B is mild diabetic.
Thus it can be seen that the value of $T_{0}$ determines whether
a person is pre-diabetic (for which $T_{0}\approx4.0$ hours) or not.
This result nearly matches with the earlier model result but more
accurate than the previous result to predict pre-diabetic condition.
Also we found the value of a important parameter $\tau$ which is
not included in the previous models. This parameter tells us how fast
glucose enters in the blood from the digestive system. This parameter
should be independent of the diabetes or blood glucose level at any
time. From Table 2 we can see that the value of $\tau$ is nearly
equal for both normal and diabetic person. So, $\tau$ is independent
of diabetes and does not depend on blood glucose level.

\begin{figure}[H]
\centering{}\includegraphics[scale=0.65]{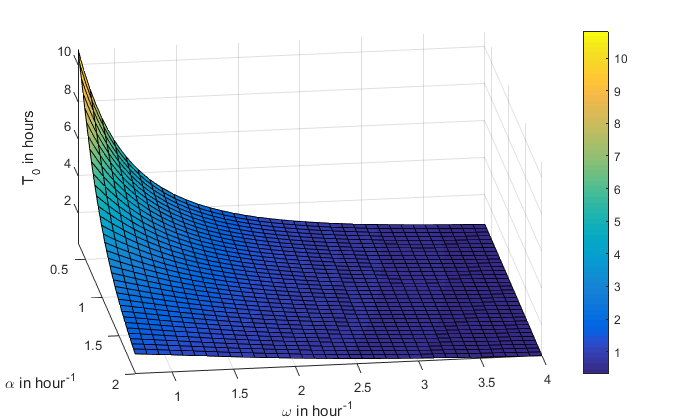}\caption{Three dimensional surface plot with different $\alpha$ and $\omega$. }
\end{figure}

\begin{figure}[H]
\centering{}\includegraphics[scale=0.65]{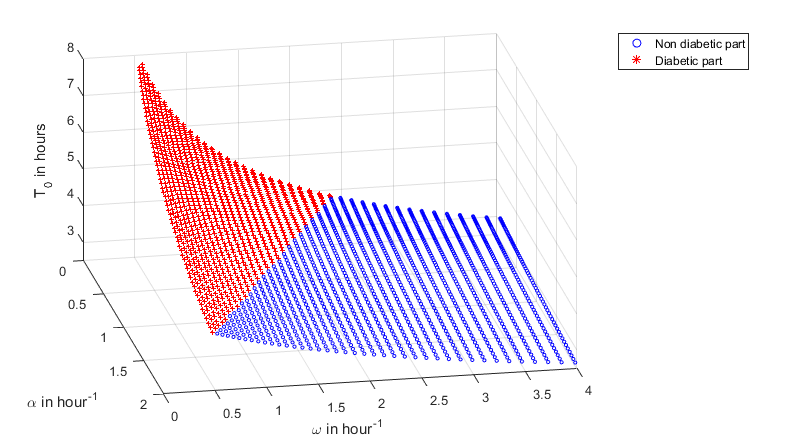}\caption{Three dimensional simulated scatter plot.}
\end{figure}

Figure 7 represents a three dimensional simulated surface plot which
shows how $T_{0}$ varies with the possible range of $\alpha$ and
$\omega$. We also see as $\alpha$ and $\omega$ decreases, $T_{0}$
increases sharply after a point. High $T_{0}$ means that blood glucose
level remains high for a long interval of time, which is the sign
of a pre-diabetic case. Figure 8 represents a three dimensional simulated
scatter plot. Here, blue dots represent normal cases (for which $T_{0}<4.0$
hours) and red dots represent pre-diabetic or diabetic (for which
$T_{0}\geq4.0$ hours) cases. This plot shows that there exists two
completely separate regions for normal and diabetic cases. It also
shows that there is a large range of $\alpha$ and $\omega$ which
can vary person to person.

\section{Discussion and Conclusion}

In this paper, we have modified an existing model to a more realistic
one by considering ingested glucose level, which is described by an
additional ordinary differential equation. Here we have assumed that
the externally ingested glucose decreases exponentially with time
by increasing the blood glucose level, which is modeled by a parameter
$\tau$ which describes how fast ingested glucose decreases. We have
found that the model is globally stable around the fixed point (0,0,0).
The solution imitates the behaviour of a damped harmonic oscillator
and it converges to basal values normally observed in the human body.
After fitting this relation to an available data-set, various fit
parameters were obtained. Using these, the value of the parameter
$T_{0}$, which is the natural time period of a damped oscillator,
was found. To conclude, we would like to point out that the most important
improvement of this model over earlier models is it's ability to predict
the vulnerability of a person to be diabetic in the future. We have
deduced that $T_{0}$ is less than 4 hours for a normal person and
for a diabetic person, this time period is greater than 4 hours, which
also matches the earlier established model predictions. Thus if a
person has this natural time period $T_{0}$ of value around 4 hours,
then it can be concluded that the person is susceptible to diabetes
in future (pre-diabetic). However, in this model we consider that
pancreas responded instantly with the blood glucose disturbance and
hormone disturbance. But a small time delay of pancreatic response
due to these disturbances will be more practical \cite{key-32}. We
can modify our model by considering this time delay. So, there is
a lot of scope for further development of this model, which can enable
precise and better control of this pre-diabetic stage and thus modify
the quality of life of a human being. 

\section*{Acknowledgment}

We would like to thank Dr. Indrani Bose and Dr. Tanaya Bhattacharyya
for their useful comments and suggestions. We also like to thank the
Department of Physics, St. Xavier's College for providing support
during this work.

\end{document}